\documentclass[pre,twocolumn,showpacs,showkeys,amsmath,amssymb]{revtex4}

\usepackage{graphicx}
\usepackage{textcomp}
\usepackage[tight]{subfigure}
\usepackage{dcolumn}
\usepackage{bm}

\begin{document}


\title{Buckling transition of nematic gels in confined geometry}

\author{Guangnan~Meng}
\altaffiliation[Present address: ]{Department of Physics \& SEAS, 9 Oxford St, Harvard University, Cambrige, Massachusetts 02138}
\email{gnmeng@seas.harvard.edu}
\affiliation{The Martin Fisher School of Physics, Brandeis University, Waltham, Massachusetts 02454-9110, USA}
\author{Robert~B.~Meyer}
\affiliation{The Martin Fisher School of Physics, Brandeis University, Waltham, Massachusetts 02454-9110, USA}%

\date{\today}

\begin{abstract}
	A spontaneous buckling transition in thin layers of monodomain nematic liquid crystalline gel was observed by polarized light microscopy. The coupling between the orientational ordering of liquid crystalline solvent and the translational ordering of crosslinked polymer backbones inside the nematic gel contributes to such buckling transition. As the nematic mesogens become more ordered when the gel is cooled down from a higher gelation temperature, the polymeric backbones tend to elongate along the direction parallel to the nematic director, which is perpendicular to the rigid glass surfaces in the experimental setup. The shape change of such confined gel sample lead to the spontaneous buckling of polymeric network and the spatial modulation of nematic liquid crystalline director, which is observed as the stripe patterns. The instability analysis was used to explain such transitions, and the relationship between the critical field, stripe's wavelength and temperature can be explained qualitatively by the rubber elasticity theory for liquid crystalline gels.
\end{abstract}

\maketitle

\section{\label{Sec:Intro}Introduction}

	Liquid crystalline gels\cite{2003LCEWarner.M;Terentjev.E} refer to special soft materials that incorporate the symmetry properties of liquid crystalline\cite{1993PhysLC_Gennes.P;Prost.J} phases into the crosslinked polymeric backbones, thus the translational response of crosslinked polymeric networks and the orientational response of liquid crystalline mesogens are coupled together. Among all the possible liquid crystalline gels, nematic gel has the simplest symmetry where the crosslinked polymeric backbones are spontaneously elongated along one certain direction (usually the nematic director $\mathbf{\hat{n}}$) under the effect of symmetry broken properties of the nematic solvent. The uniaxial prolate ellipsoidal polymer backbones can be described by a step length tensor $l_{ij}=l_\perp\delta_{ij}+(l_\parallel-l_\perp)n_in_j$  and the anisotropic parameter $r$ is defined as the ratio of effective step length of polymer coil parallel ($l_{\parallel}$) and perpendicular ($l_{\perp}$) to the nematic director $\mathbf{\hat{n}}$. The value of $r$ depends on the symmetry properties of the nematic solvent: $r$ increases as the system becomes more ordered, i.e. lower temperature for thermotropic liquid crystal. When there are no rigid mechanical constrains, such relationship can be verified by observing the macroscopic shape change of the nematic gel between the isotropic phase and the nematic phase\cite{__In-Preparation_Vol:_Pg:_Meng.G;Meyer.R,2002_11_Physical-Review-Letters_Vol:89_Pg:225701_Selinger.J;Jeon.H;etal}. As the temperature decreases, the mono-domain nematic gel sample will become elongated along the direction of nematic director when sample changes from the isotropic phase into the nematic phase. When the material is confined by rigid boundaries, for example, on the direction of elongation during the cooling, a buckling transition is expected to happen and it has been experimentally observed as stripe patterns under polarized light microscopy\cite{2006_04_Physical-Review-Letters_Vol:96_Pg:147802_Verduzco.R;Meng.G;etal}. Here, we report another buckling transition in thin layers of same nematic liquid crystalline gels within different confined geometries. The physical reasons for such buckling transition can be qualitatively interpreted by the coupling between the mechanical response of crosslinked polymeric backbones and the orientation response of the nematic solvent. The instability analysis were applied to explain the experimental phenomena such as the temperature dependence of critical point and wavelength of periodic patterns. The study about this buckling phenomena is helpful to provide insights to the buckling transition found in other soft materials, i.e. microtubules\cite{1996_Physical-Review-Letters_Vol.76_No.21_Pg.4078-4081_Elbaum.M;Fygenson.D;Libchaber.A_}, F-actin networks\cite{2007_Nature_Vol.445_No.7125_Pg.295-298_Chaudhuri.O;Parekh.S;Fletcher.D_}. 

\section{Material and Experimental}

	Nematic gel material was synthesized in Kornfield's group\cite{2004_11_Macromolecules_Vol:37_Pg:8730--8738_Kempe.M;Kornfield.J;etal,2004_05_Macromolecules_Vol:37_Pg:3569--3575_Kempe.M;Kornfield.J;etal,2004_03_Nature-Materials_Vol:3_Pg:139--140_Palffy-Muhoray.P;Meyer.R,2004_03_Nature-Materials_Vol:3_Pg:177--182_Kempe.M;Scruggs.N;etal}. Briefly, 5-wt\% of ABA triblock copolymer, which consist of polystyrene as end blocks and side group liquid crystalline polymer as middle blocks, were dissolved into a nematic solvent (4-\emph{n}-pentyl-4'-cyanobiphenyl, 5CB). The formation of weak physical network is controlled by the order parameter of the solvent: the polystyrene end blocks are soluble in the isotropic phase and aggregate in the nematic phase. The phase transition temperature of such nematic gel ($T_{\mathrm{NI}}\approx 37$\textcelsius ) is very close to the transition temperature  of 5CB ($T_{\mathrm{NI}}\approx 35$\textcelsius ), and the reversibility of the physical crosslinking mechanism allows repeatable experiments being easily conducted on the same sample.

\begin{figure}
 \def \picwidth {0.45\textwidth}
 \includegraphics[width=\picwidth]{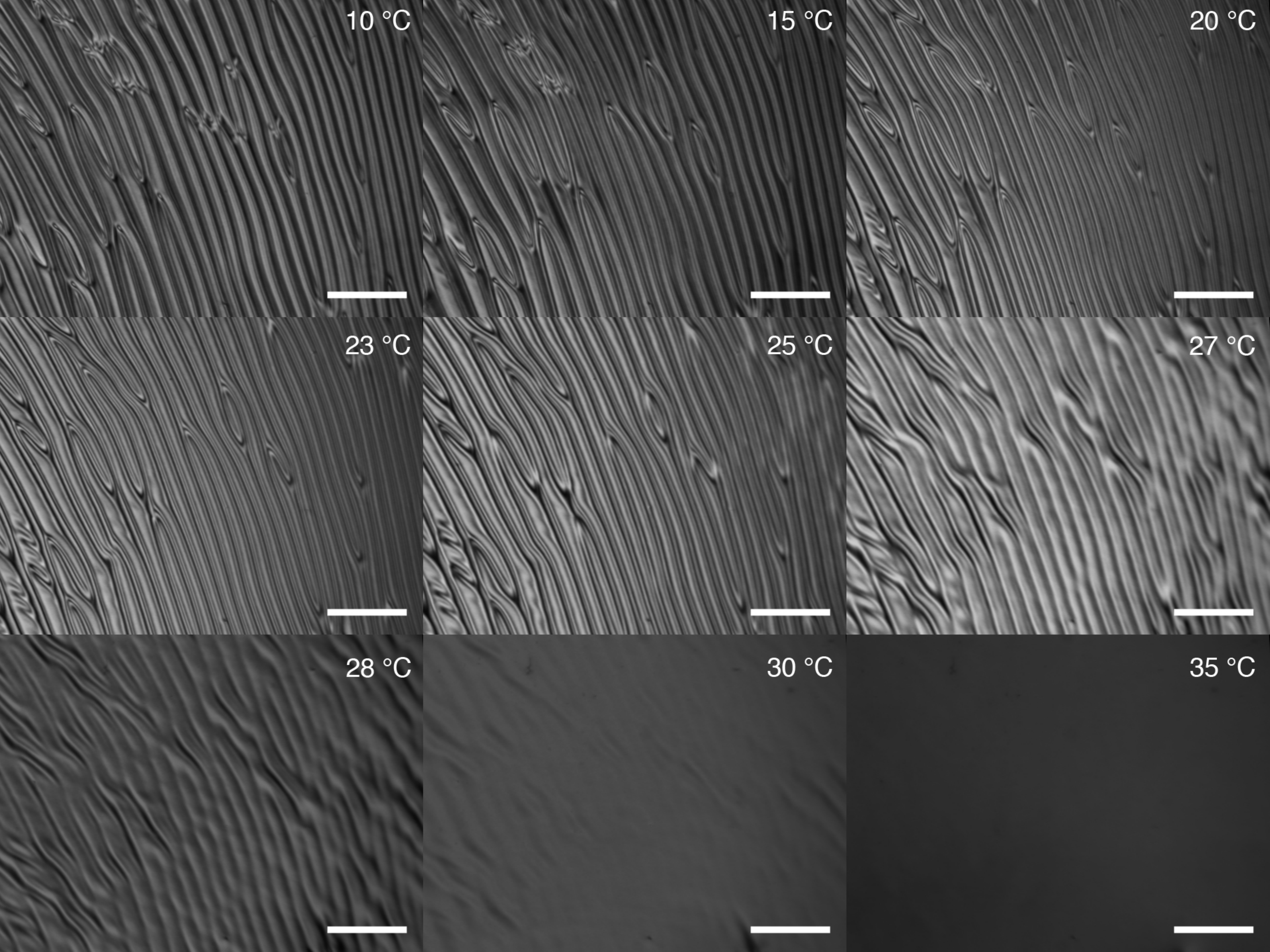}
 \caption{\label{Fig:BucklingTransition}
    	Optical micrographs of birefringent stripe pattern of buckled nematic gel observed through polarized optical microscopy with applied electric field decreased to zero at different temperatures.
	The scaling bars stand for 50$\mu\mathrm{m}$ in all the images.
    }
  \end{figure}

\begin{figure}
 \def \picwidth {0.225\textwidth} 
   \subfigure[
   	]{\label{Fig:WavelengthTemp}\includegraphics[width=\picwidth]{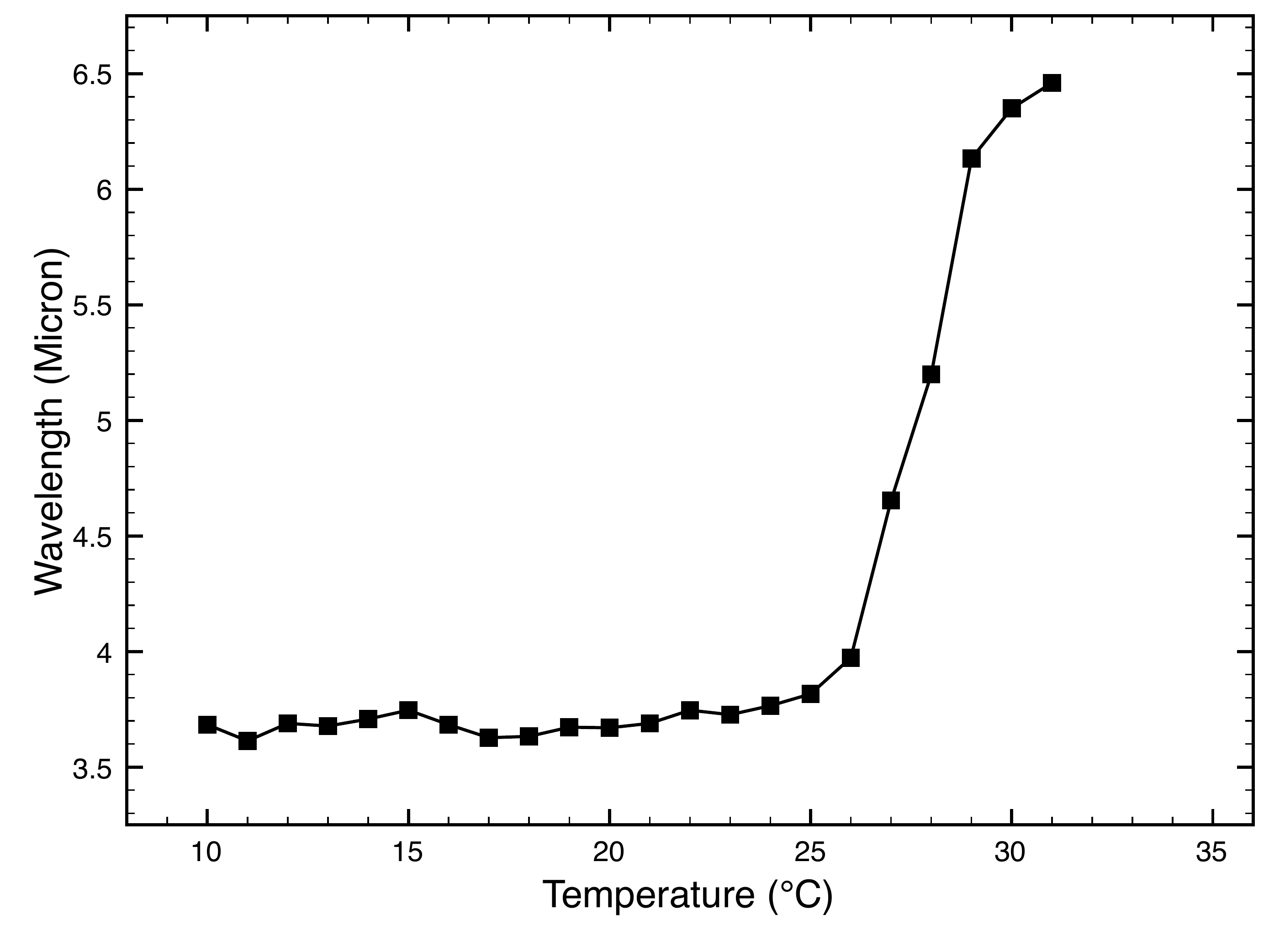}}
   \subfigure[
   	]{\label{Fig:FieldTemp}\includegraphics[width=\picwidth]{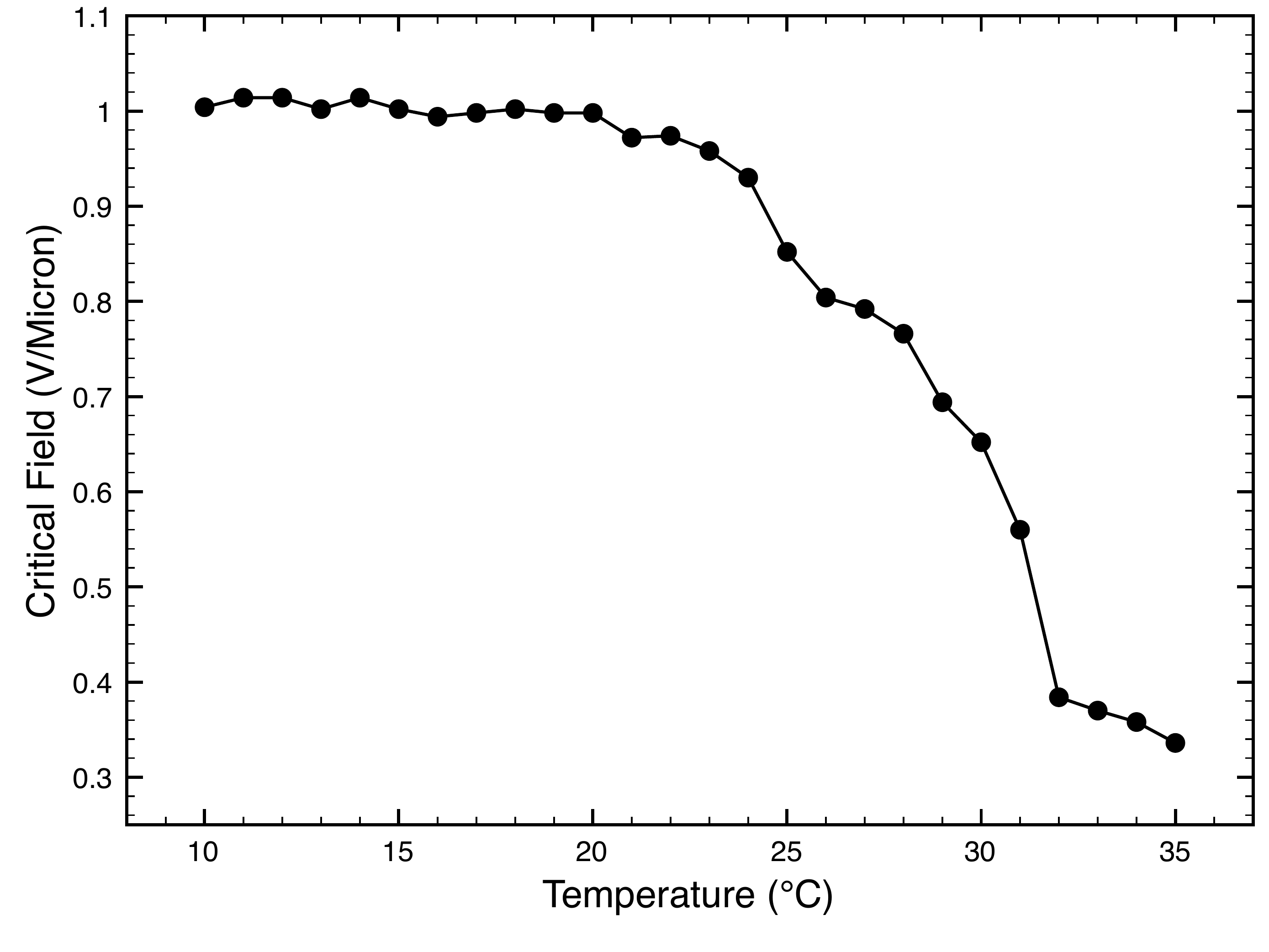}}
   \caption{\label{Fig:TempExperiment}
      	Experimental measurement about the temperature dependence of \subref{Fig:WavelengthTemp} wavelength of stripes in the buckled state and \subref{Fig:FieldTemp} critical field ($E_{\mathrm{C}}$) of the buckling transition in nematic gel.
    }
  \end{figure}

	The nematic gel is loaded in a 25$\mu\mathrm{m}$ thick homeotropic electro-opitcal cell, in which a thin layer of \emph{n},\emph{n}-dimethyl-\emph{n}-octadecyl-3-aminopropyl-trimethoxysilyl-chloride 
were spin coated on the surface of transparent indium-tin oxide conductors of glass slides. In the presence of a strong applied electric field  ($E_{0}=3\mathrm{V}/\mu\mathrm{m}$, 1kHz) across the cell, the nematic mesogens can be easily aligned vertically throughout the cell, where the long axis pointing perpendicularly to the boundary surfaces. The temperature of the sample was controlled by a peltier-based microscope stage during the observation. Initially, the sample was heated up to 45\textcelsius\ in the isotropic phase with no crosslinked polymer networks. The sample was cooled down (2\textcelsius/minute) across its $T_{\mathrm{NI}}$ to certain final temperatures ($T_{f}$) and  a thin layer film of mono-domain nematic gel was obtained during the gelation throughout the cell volume, in which $\mathbf{\hat{n}}$ points perpendicularly to the boundary surfaces. The sample appeared homogenous dark under the cross polarized optical microscopy while the aligning field maintained its original magnitude. When the electric filed was turned off, birefringent stripe patterns with wavelength about 5$\mu\textrm{m}$ appeared throughout the sample, as shown in Fig.~\ref{Fig:BucklingTransition}. Both the wavelength of stripe pattern and the critical field ($E_{\mathrm{C}}$), at which the sample changes from homogeneous dark to birefringent patterned, depend on the sample's final cooling temperature ($T_{f}$), such temperature dependance are recorded and plotted in Fig.~\ref{Fig:TempExperiment}. It can be seen that  both $E_{\mathrm{C}}$ and wavelength stay in a plateau when 10\textcelsius$<T_{f}<$24\textcelsius\ , and when $T_{f}>24$\textcelsius\  $E_{\mathrm{C}}$ decreases as  $T_{f}$ increases, while the wavelength increases.

 \section{Discussions}
  
 \begin{figure}
 \def \picwidth {0.225\textwidth}
  \subfigure[ $T_{i}\lessapprox T_{\mathrm{NI}}$]{\label{Fig:BucklingInitial}\includegraphics[width=\picwidth]{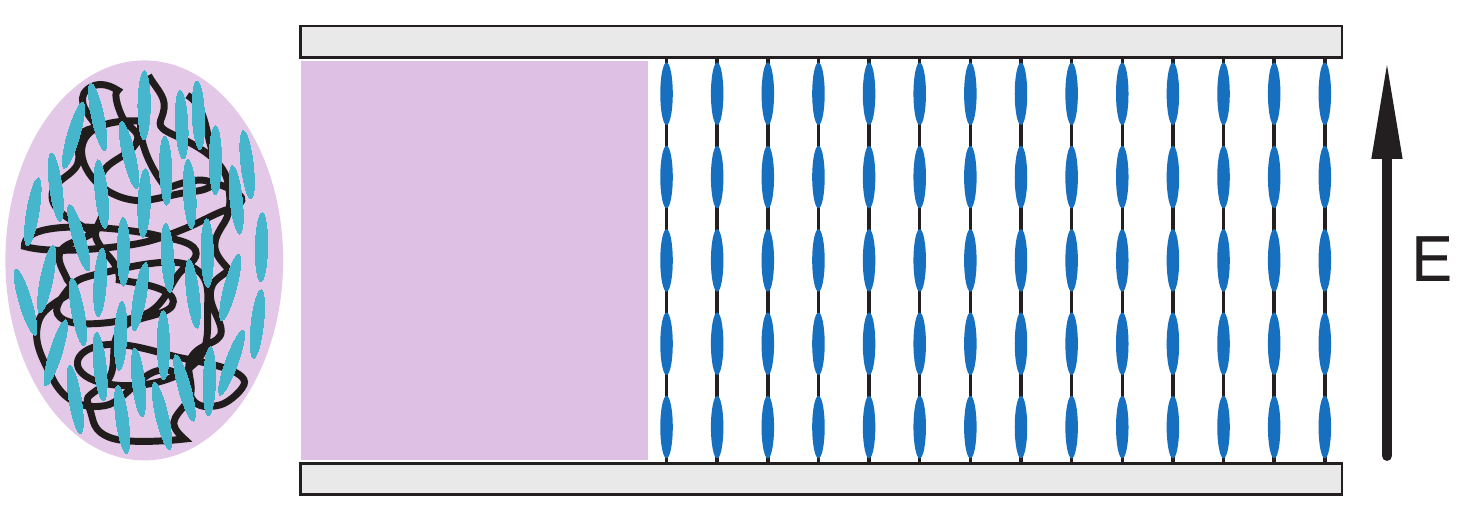}}
  \subfigure[ $T_{f}<T_{i}$]{\label{Fig:BucklingFinal}\includegraphics[width=\picwidth]{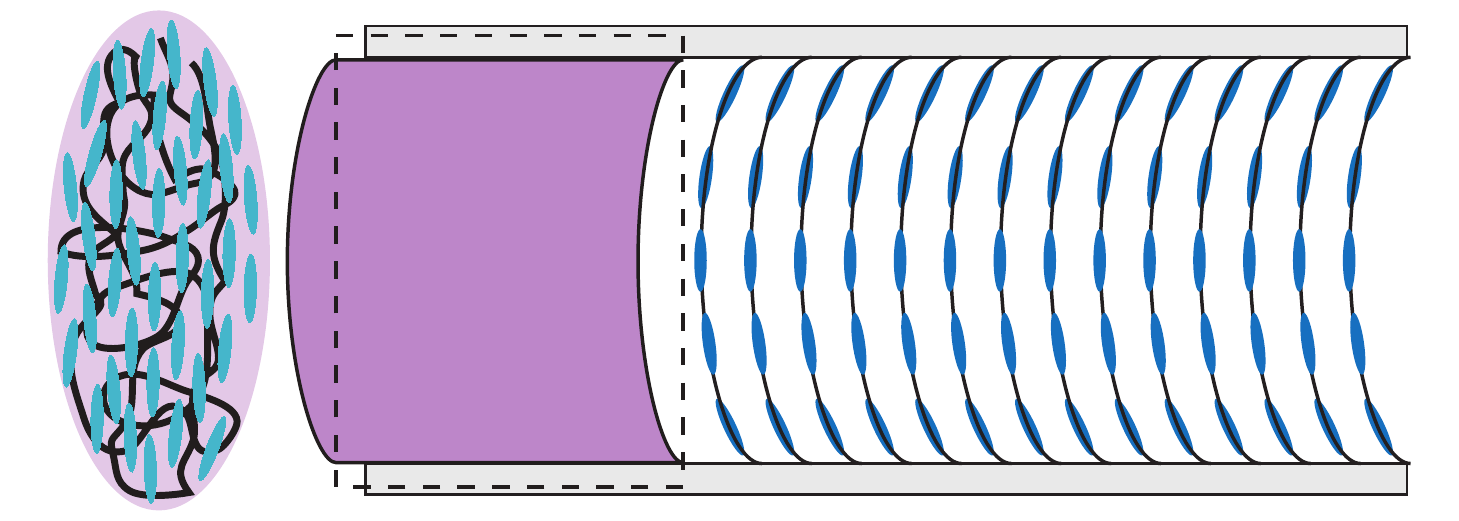}}
  \caption{\label{Fig:BucklingDiagram}
   	Diagrams of the buckling transition in nematic gel. The rigid boundaries cause the material buckling within itself as the polymeric backbones elongate in a more ordered state  \subref{Fig:BucklingFinal}  in comparison to a less ordered gelation state \subref{Fig:BucklingInitial}, at which the liquid crystal mesogens are aligned vertically by the applied electric field $\mathbf{E}$.
  }
 \end{figure}

    The driving force of such transition can be attributed to the thermo-mechanical-optical coupling between the nematic liquid crystalline solvent and anisotropic crosslinked polymeric backbones in the nematic gel within a confined boundary condition. Diagrams in Fig.~\ref{Fig:BucklingDiagram} are used to interpret the physical reasons for such transition. Initially, the monodomain nematic gel is formed at a higher initial temperature ($T_{i}\lessapprox T_{\mathrm{NI}}$) within the glass cell as the electric field is applied across the cell, as show in Fig.~\ref{Fig:BucklingInitial}. The micro picture of the nematic gel is sketched as the ellipsoid and macroscopic shape of the gel is expressed as the square shape in the diagram. As the temperature is lowered to final temperature ($T_{f}<T_{i}$), the system becomes more ordered and the anisotropic polymeric coil will elongate along the nematic director's direction, as shown in the ellipsoid of Fig.~\ref{Fig:BucklingFinal}. If there is no boundaries to confine the shape of the material, the macroscopic shape of the mono-domain gel sample will elongate vertically, which is illustrated as the rectangular dashed line. Such ``artificial muscle'' effect has been observed experimentally in nematic elastomers and gels\cite{2002_11_Physical-Review-Letters_Vol:89_Pg:225701_Selinger.J;Jeon.H;etal}. When the sample is put into an environment with rigid constrains, e.g. a cell with two glass slides glued together, the material has to buckle within the boundaries in order to gain the elongation along the direction perpendicular to the rigid boundaries. Due to the coupling between the translational response of the polymeric coil and rotational response of the liquid crystalline mesogen, the nematic director $\hat{\mathbf{n}}$ will rotate correspondingly. The applied electric field with enough magnitude can be used to keep the nematic director aligned vertically, and the material will buckle spontaneously as the electric field is decreased due to the instability behavior. To prove our physical explanation, detailed analytical calculations are conducted and discussed in the following.

\subsection{Modeling and Free Energy Calculation}
	The coordinate origin is selected at the middle of the cell gap and the $z$-axis is perpendicular to the cell boundaries ($z=\pm d/2$). Initially, the aligning electric field is applied along the $z$-axis ($\mathbf{E}=E_{0}\hat{\mathbf{z}}$) and the nematic director is initially aligned vertically ($\mathbf{\hat{n}}^0=\hat{\mathbf{z}}$) as well, as shown in Fig.~\ref{Fig:BucklingInitial}. The superscript $0$ is added onto parameters for the initial gelation state of the material. The polymeric networks are formed during the gelation process at temperature $T_{i}$, with $r^{0}$ as the anisotropic parameter. When the material is cooled to a lower final temperature $T_{f}<T_{i}$, the crosslinked polymeric network will be more elongated along nematic director $\hat{\mathbf{n}}$ as the nematic solvent become more ordered and the anisotropy parameter $r$ that is larger than the initial value, $r>r^0$. When the electric field is decreased to zero ($\mathbf{E}=0$), the elongation of the polymeric coils within the nematic liquid crystalline gel buckle with a displacement field within the $xz$-plane: $\mathbf{R}=\zeta\cos{kx}\cos{qz}\hat{\mathbf{x}}+\eta\sin{kx}\sin{qz}\hat{\mathbf{z}}$, in which $k$ is the wavevector of the shear wave within $xy$-plane, and we select it along the $x$-axis, $q$ is the wavevector along the $z$-direction. The value of $q$ is determined by the sample thickness $d$ as $q=\pi/d$ for the first harmonic mode. $\zeta$ and $\eta$ are related by the incompressibility condition\cite{2003LCEWarner.M;Terentjev.E} as $q\eta=k\zeta$. Such shear motion of the polymeric network will induce the nematic director to rotate within $xz$-plane of small amplitude $\xi$ about $z$-axis: $\mathbf{\hat{n}}=\xi\cos{kx}\sin{qz}\hat{\mathbf{x}}+\hat{\mathbf{z}}$. 

	We plug these conditions into the formulas for free energy density, which include the Frank curvature elastic energy of the nematic solvent\cite{1993PhysLC_Gennes.P;Prost.J} and the nematic rubber elastic energ of crosslinked polymeric backbones\cite{2003LCEWarner.M;Terentjev.E}:
\begin{widetext}
\begin{eqnarray}\label{Eq:NematicCurvature}
	F&=&\frac{1}{2}\left( K_{S}\left(\nabla\cdot\mathbf{\hat{n}}\right)^2+K_{B}\left(\mathbf{\hat{n}}\times\left(\nabla\times\mathbf{\hat{n}}\right)\right)^2\right)-\frac{1}{2}\epsilon_a(\mathbf{E\cdot\hat{n}})^2\nonumber\\
	&&+\frac{1}{2}\mu\cdot Tr(\underline{\underline{l^0}}\cdot\underline{\underline{\Lambda}}^T\cdot\underline{\underline{l}}^{-1}\cdot	\underline{\underline{\Lambda}})-\frac{1}{2}A\mu\cdot Tr(\underline{\underline{\Lambda}}^T\cdot\underline{n}\cdot\underline{\underline{\Lambda}}-\underline{n^0}\cdot\underline{\underline{\Lambda}}^T\cdot\underline{n}\cdot	\underline{\underline{\Lambda}}).
\end{eqnarray}
\end{widetext}
	In Eq.~\ref{Eq:NematicCurvature}, $K_{S}$ and $K_{B}$ are the curvature elastic constants for the nematic solvent; $\mu$ is the shear modulus of the gel and $A$ is the semisoftness coefficient; $\underline{\underline{\Lambda}}$ is the Cauchy strain tensor, $\Lambda_{ij}=\delta_{ij}+\partial R_{i}/\partial x_{j}$.

By averaging this free energy density over the space, minimizing with respect to $\xi$, and keeping only terms only to second order in $\zeta$, the free energy density $f$ in the material can be written in Eq.~\ref{Eq:FinalFreeEnergy}.
\begin{widetext}
  \begin{eqnarray}
  \label{Eq:FinalFreeEnergy}
    f&=&\frac{\mu\zeta^2}{4 q^2r\big(1-r^{0}+r(A-1+\frac{\epsilon_aE^2+K_{S}k^2+K_{B}q^2}{\mu}+r^{0})\big)}\times\nonumber\\
    &&\bigg(k^6r(1+Ar)\frac{K_{S}}{\mu} +q^4r^{0}\big(r+(A-1+\frac{\epsilon_aE^2+K_{B}q^2}{\mu})r^2+(r-1)r^{0}\big) \nonumber\\
    &&+k^4r\big((1+Ar)(1+r^{0})-r-Ar^{0}+(1+Ar)\frac{\epsilon_aE^2+K_{B}q^2}{\mu}+(r+r^{o})\frac{K_{S}q^2}{\mu}-\frac{r^{0}}{r}\big)\nonumber\\
    &&+k^2q^2\Big((3-r^{0})r^{0}+r\big(1+r^{0}(3A-6+\frac{\epsilon_aE^2+K_{B}q^2}{\mu}+r^{0})\big)\Big)\nonumber\\
    &&+r^2\big(A-1+\frac{\epsilon_aE^2}{\mu}+3r^{0}-2Ar^{0}+\frac{q^2}{\mu}(K_{B}+K_{S}r^{0})\big)\bigg) +o(\zeta^4).
  \end{eqnarray}
  \end{widetext}
	It can be seen that the free energy is proportional to the square of the perturbation's amplitude $\zeta^2$. If $f<0$, the perturbed state would be stable and the system would have a transition with $\zeta\ne0$; on the other hand, if $f>0$, the perturbed state is unstable and the system would stay in its initial state with $\zeta=0$. $f=0$ is the critical point at which the transition starts. In this way, the values of the material's physical parameters determine the instability behavior under different external experimental conditions. e.g. applied field, temperature. 
	We can plug the known parameters of the nematic liquid crystalline gel\cite{2004_03_Nature-Materials_Vol:3_Pg:177--182_Kempe.M;Scruggs.N;etal,2006_04_Physical-Review-Letters_Vol:96_Pg:147802_Verduzco.R;Meng.G;etal} into Eq.~\ref{Eq:FinalFreeEnergy}, as $d=25 \mu\mathrm{m}$, $\mu=50\mathrm{Jm^{-3}}$, $A=0.1$, $K_{B}=10^{-11}\mathrm{Jm^{-1}}$, $K_{S}=1.5\times10^{-11}\mathrm{Jm^{-1}}$, $\epsilon_a=15\epsilon_0$. We choose $r^0=1.5$ for the initial gelation state at temperature $T_{i}$,  and $r$ depends on the order parameter of the nematic solvent in the gel for specific temperature $T_{f}$ with elongated polymeric coil.

 \subsection{Instability Analysis: Critical Field}

\begin{figure}
 \def \picwidth {0.225\textwidth}
 \centering
 \subfigure[ $f_{k\to0}$ vs. $E$]{\label{Fig:FreeEnergyField}\includegraphics[width=\picwidth]{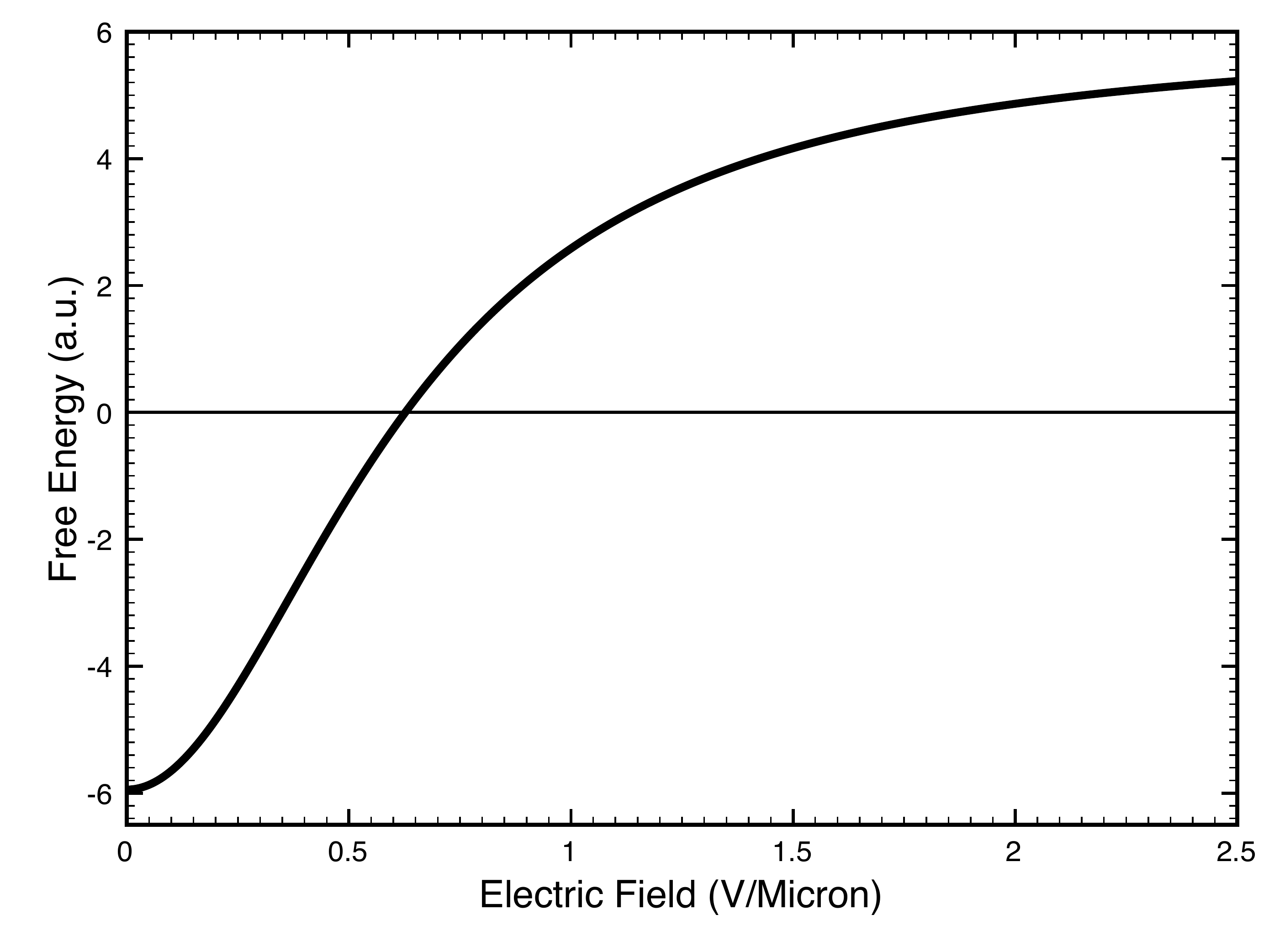}}
 \subfigure[ $E_{\mathrm{C}}$ vs. $r$]{\label{Fig:FieldRPlot}\includegraphics[width=\picwidth]{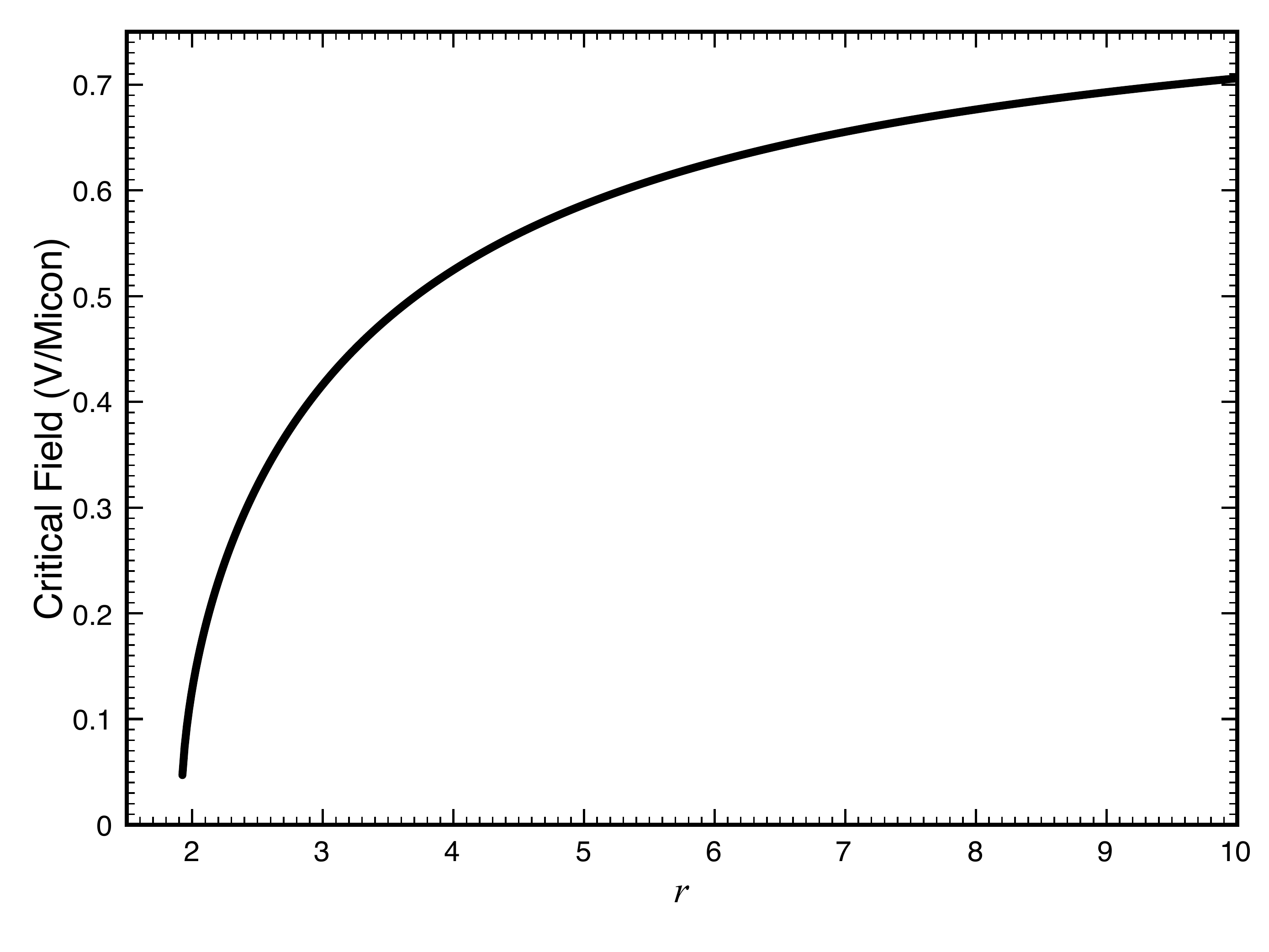}}
 \subfigure[ $f_{E\to0}$ vs. $k$]{\label{Fig:FreeEnergyK}\includegraphics[width=\picwidth]{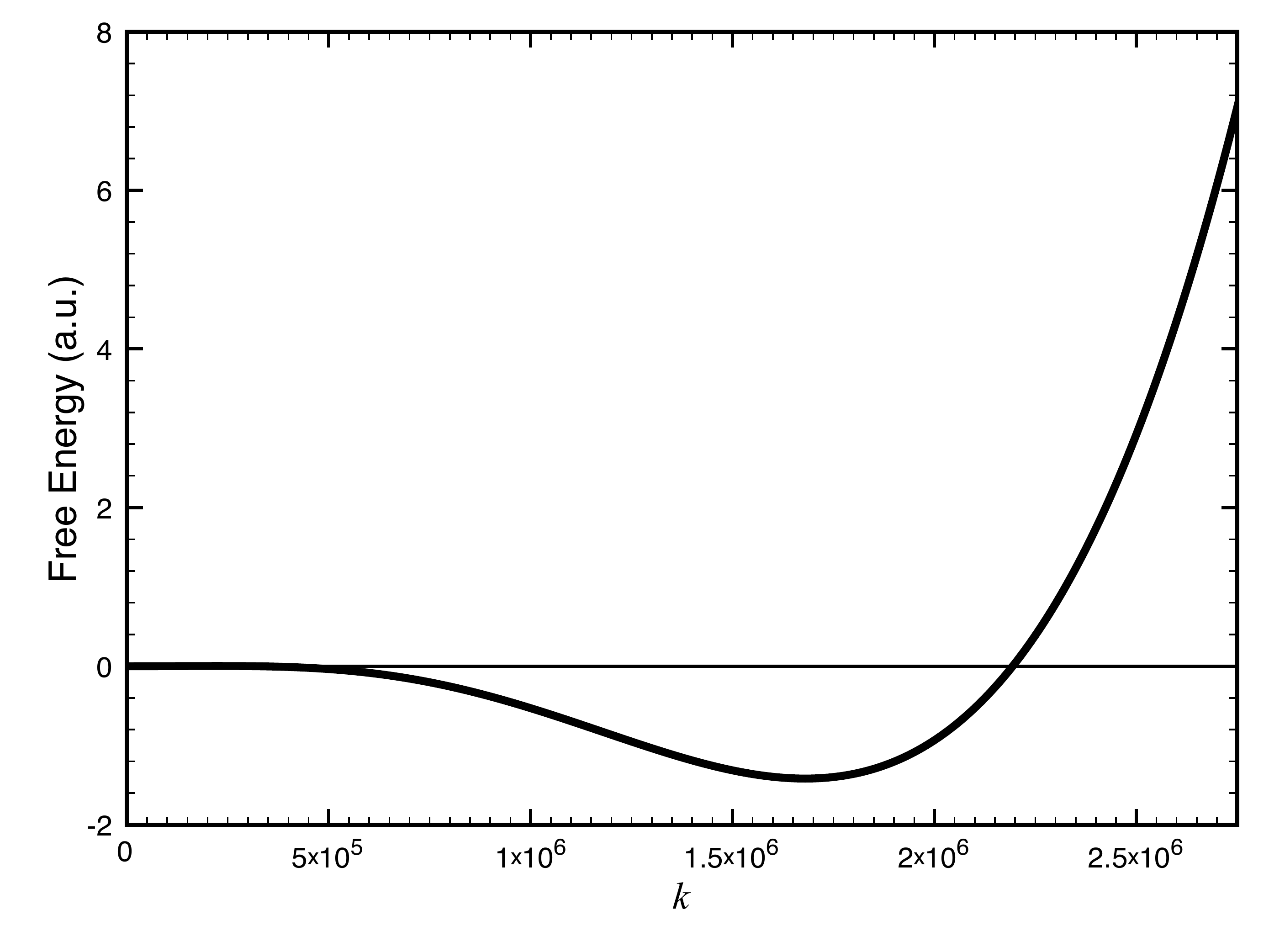}}
 \subfigure[ $\lambda$ vs. $r$]{\label{Fig:KR}\includegraphics[width=\picwidth]{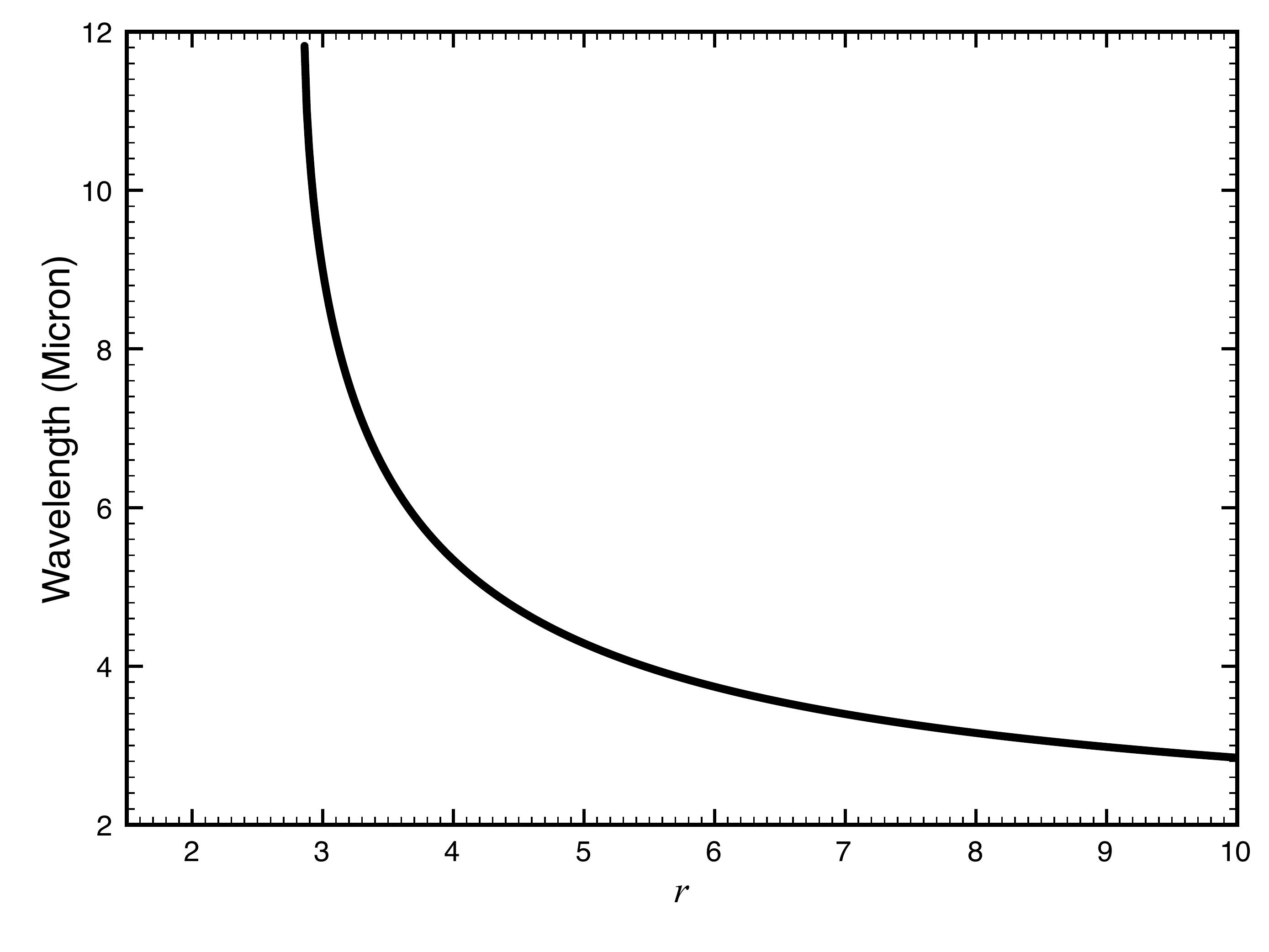}}
 \caption{
 	The free energy density ($f$) depends both on \subref{Fig:FreeEnergyField} electric field and \subref{Fig:FreeEnergyK} wavevector $k$. The free energy is minimized by decreasing applied field to zero and spatial modulation of director field $\hat{\mathbf{n}}$ with a finite $k$. Both critical field $E_{\mathrm{C}}$ and wavelength depend on the anisotropic properties of final buckled state: as the nematic gel is more ordered at its final state (corresponding lower $T_{f}$), \subref{Fig:FieldRPlot} $E_{\mathrm{C}}$ increases and \subref{Fig:KR} wavelength decreases.
    }
  \end{figure}

 	For the case of a homogeneous birefringence change as the basic mode of transition, $f$ can be further simplified by setting $k=0$: 
 \begin{widetext}
 \begin{equation}\label{Eq:EnergyField}
  	f_{k\to0}=\frac{\mu \zeta^2q^2r^0}{4r}
  		\left(\frac{r+(r-1)r^0+r^2\left(A-1+\frac{\epsilon_{a}E^2+K_{B}q^2}{\mu}\right)}{1-r^0+r\left(A-1+r^0+\frac{\epsilon_{a}E^2+K_{B}q^2}{\mu}\right)}\right).
 \end{equation}
 \end{widetext}
	Fig.~\ref{Fig:FreeEnergyField} shows the plot of $f_{k\to0}$ as a function of electric field's intensity $E$,  which corresponds to the situation when we decrease the electric field at a certain temperature of final state. The free energy is positive when the electric field is still large enough to keep the vertical alignment of the nematic directors across the cell, the buckling transition is energetic unfavorable; as the electric field is further decreased, the energy become negative, which means the system transforms into the buckled state to minimize the free energy. The critical electric field $E_{\mathrm{C}}$ can be found at the point where the free energy equals zero. The relationship between the $E_C$ and $r$ can be studied numerically and plotted in Fig.~\ref{Fig:FieldRPlot}, where $E_{\mathrm{C}}$ increases with $r$, in another word, $E_{\mathrm{C}}$ decreases with final temperature $T_{f}$. This agree qualitatively with our experimental measurement in Fig.~\ref{Fig:FieldTemp}: when $r=3.5$, critical field is calculated as $E_{\mathrm{C}}=0.48\textrm{V}/\mu\mathrm{m}$ comparing with experimental value $0.56\textrm{V}/\mu\mathrm{m}$ at 31\textcelsius ; when $r=5.8$ corresponding to a lower $T_{f}$, critical field is calculated as $E_{\mathrm{C}}=0.621\textrm{V}/\mu\mathrm{m}$ comparing with experimental value $0.852\textrm{V}/\mu\mathrm{m}$ at 25\textcelsius .

 \subsection{Instability Analysis: Stripe Wavelength}
 	Since the nematic gel buckles when the applied electric field removed ($E=0$), the free energy $f$ depends on the wavevector $k$ in $xy$-plane:
 \begin{widetext}
 \begin{eqnarray}\label{Eq:EnergyK}
    f_{E\to0}&=&\frac{\mu\zeta^2}{4 q^2r\big(1-r^{0}+r(A-1+\frac{K_{S}k^2+K_{B}q^2}{\mu}+r^{0})\big)}\times\bigg(k^6K_{S}r\frac{1+Ar}{\mu} \nonumber \\
    &&+k^4r\big(1+r^{0}-Ar^{0}+\frac{q^2}{\mu}(K_{B}+K_{S}r^{0})\big)+k^2q^2\Big((3-r^{0})r^{0}+r\big(1+r^{0}(3A-6+\frac{K_{B}q^2}{\mu}+r^{0})\big)\Big) \nonumber \\
    &&+q^4r^{0}\big(r+(A-1+\frac{K_{B}q^2}{\mu})r^2+(r-1)r^{0}\big)+r^2\big(A-1+3r^{0}-2Ar^{0}+\frac{q^2}{\mu}(K_{B}+K_{S}r^{0})\big)\nonumber \\
    &&+r^2\big(\frac{K_{S}q^2}{\mu}-1+A(1+\frac{K_{B}q^2r^{0}}{\mu})\big)-r^{0}\bigg).
  \end{eqnarray}
  \end{widetext}
  	Fig.~\ref{Fig:FreeEnergyK} shows the plot of $f_{E\to0}$ as a function of wavevectore $k$. It can be seen that free energy is negative at a finite $k$, which corresponding to the stripe pattern observed experimetally. The spatial modulation of the translational order ($\mathbf{R}$) and the orientational order ($\hat{\mathbf{n}}$) further minimize the free energy, which is similar to the stripe pattern observed in previous experimental observation in the planar aligned sample\cite{2006_04_Physical-Review-Letters_Vol:96_Pg:147802_Verduzco.R;Meng.G;etal}. Furthermore, wavelength can be numerically calculated as a function of the sample's anisotropic parameter $r$ at final state, which is plotted in Fig.~\ref{Fig:KR}. It can be seen that wavelength decreases as $r$ increases, in another word, stripe's wavelength is smaller for lower final temperature $T_{f}$. This agrees with our experimental observations in Fig.~\ref{Fig:WavelengthTemp}: when $r=3.5$, the wavelength is calculated as $6.48\mu\mathrm{m}$ comparing with experimental value $6.46\mu\mathrm{m}$ at 31\textcelsius ; when $r=5.8$ corresponding to a lower $T_{f}$, the wavelength is calculated as $3.82\mu\mathrm{m}$ comparing with experimental value $3.82\mu\mathrm{m}$ at 25\textcelsius .
  
 
 	Currently, the analytic relationship between temperature $T$ (or the order parameter of nematic liquid crystals) and the anisotropy parameter $r$ of polymeric networks is not known experimentally. Therefore we can not fit our experimental measurement of the critical field and wavelength with temperature. The theoretical calculation can only be compared with experimental data qualitatively. It can be seen that it agrees well qualitatively with the experimental measurements.
 
\section{\label{Sec:Conclusion}Conclusions}

	In summary, the spontaneous buckling transitions of thin layers of nematic liquid crystalline gel in a homeotropic cell were observed by polarized light microscopy. This is good example to show the coupling between the liquid crystalline ordering and the crosslinked polymer backbones inside the nematic gel material. As the nematic mesogens become more ordered when the gel is cooled down from the initial crosslinking stage with a higher temperature, the polymer network tends to elongate along the direction parallel to the initial nematic director, which is perpendicular to the rigid glass surfaces in the experimental setup. The shape change of such confined gel sample lead to the spontaneous buckling transition. The applied electric field will change the in stability behavior, and spatial modulated stripe pattern in orientational ordering of nematic solvent helps to accommodate the buckling transformation of gel network and minimize the free energy. The experimental observation and measurement can be can be explained qualitatively at different temperature.

\section{Acknowledgments}

\begin{acknowledgments}
	We gratefully thank for Julia~A.~Kornfield' group at California Institute of Technology for providing nematic gel material. This research was supported by NSF Grant No. DMR-0322530.
\end{acknowledgments}

\bibliographystyle{apsrev}

\begin{thebibliography}{11}
\expandafter\ifx\csname natexlab\endcsname\relax\def\natexlab#1{#1}\fi
\expandafter\ifx\csname bibnamefont\endcsname\relax
  \def\bibnamefont#1{#1}\fi
\expandafter\ifx\csname bibfnamefont\endcsname\relax
  \def\bibfnamefont#1{#1}\fi
\expandafter\ifx\csname citenamefont\endcsname\relax
  \def\citenamefont#1{#1}\fi
\expandafter\ifx\csname url\endcsname\relax
  \def\url#1{\texttt{#1}}\fi
\expandafter\ifx\csname urlprefix\endcsname\relax\def\urlprefix{URL }\fi
\providecommand{\bibinfo}[2]{#2}
\providecommand{\eprint}[2][]{\url{#2}}

\bibitem[{\citenamefont{Warner and
  Terentjev}(2003)}]{2003LCEWarner.M;Terentjev.E}
\bibinfo{author}{\bibfnamefont{M.}~\bibnamefont{Warner}} \bibnamefont{and}
  \bibinfo{author}{\bibfnamefont{E.~M.} \bibnamefont{Terentjev}},
  \emph{\bibinfo{title}{Liquid Crystal Elastomers}} (\bibinfo{publisher}{Oxford
  University Press}, \bibinfo{year}{2003}).

\bibitem[{\citenamefont{de~Gennes and
  Prost}(1993)}]{1993PhysLC_Gennes.P;Prost.J}
\bibinfo{author}{\bibfnamefont{P.~G.} \bibnamefont{de~Gennes}}
  \bibnamefont{and} \bibinfo{author}{\bibfnamefont{J.}~\bibnamefont{Prost}},
  \emph{\bibinfo{title}{The Physics of Liquid Crystals}}
  (\bibinfo{publisher}{Oxford University Press}, \bibinfo{year}{1993}),
  \bibinfo{edition}{2nd} ed.

\bibitem[{\citenamefont{Meng and
  Meyer}()}]{__In-Preparation_Vol:_Pg:_Meng.G;Meyer.R}
\bibinfo{author}{\bibfnamefont{G.}~\bibnamefont{Meng}} \bibnamefont{and}
  \bibinfo{author}{\bibfnamefont{R.~B.} \bibnamefont{Meyer}}, \bibinfo{note}{in
  Preparation}.

\bibitem[{\citenamefont{Selinger et~al.}(2002)\citenamefont{Selinger, Jeon, and
  Ratna}}]{2002_11_Physical-Review-Letters_Vol:89_Pg:225701_Selinger.J;Jeon.H;%
etal}
\bibinfo{author}{\bibfnamefont{J.~V.} \bibnamefont{Selinger}},
  \bibinfo{author}{\bibfnamefont{H.~G.} \bibnamefont{Jeon}}, \bibnamefont{and}
  \bibinfo{author}{\bibfnamefont{B.~R.} \bibnamefont{Ratna}},
  \bibinfo{journal}{Physical Review Letters} \textbf{\bibinfo{volume}{89}},
  \bibinfo{pages}{225701} (\bibinfo{year}{2002}).

\bibitem[{\citenamefont{Verduzco et~al.}(2006)\citenamefont{Verduzco, Meng,
  Kornfield, and
  Meyer}}]{2006_04_Physical-Review-Letters_Vol:96_Pg:147802_Verduzco.R;Meng.G;%
etal}
\bibinfo{author}{\bibfnamefont{R.}~\bibnamefont{Verduzco}},
  \bibinfo{author}{\bibfnamefont{G.}~\bibnamefont{Meng}},
  \bibinfo{author}{\bibfnamefont{J.~A.} \bibnamefont{Kornfield}},
  \bibnamefont{and} \bibinfo{author}{\bibfnamefont{R.~B.} \bibnamefont{Meyer}},
  \bibinfo{journal}{Physical Review Letters} \textbf{\bibinfo{volume}{96}},
  \bibinfo{pages}{147802} (\bibinfo{year}{2006}).

\bibitem[{\citenamefont{Elbaum et~al.}(1996)\citenamefont{Elbaum, Fygenson, and
  Libchaber}}]{1996_Physical-Review-Letters_Vol.76_No.21_Pg.4078-4081_Elbaum.M%
;Fygenson.D;Libchaber.A_}
\bibinfo{author}{\bibfnamefont{M.}~\bibnamefont{Elbaum}},
  \bibinfo{author}{\bibfnamefont{D.}~\bibnamefont{Fygenson}}, \bibnamefont{and}
  \bibinfo{author}{\bibfnamefont{A.}~\bibnamefont{Libchaber}},
  \bibinfo{journal}{Physical Review Letters} \textbf{\bibinfo{volume}{76}},
  \bibinfo{pages}{4078} (\bibinfo{year}{1996}).

\bibitem[{\citenamefont{Chaudhuri et~al.}(2007)\citenamefont{Chaudhuri, Parekh,
  and
  Fletcher}}]{2007_Nature_Vol.445_No.7125_Pg.295-298_Chaudhuri.O;Parekh.S;Flet%
cher.D_}
\bibinfo{author}{\bibfnamefont{O.}~\bibnamefont{Chaudhuri}},
  \bibinfo{author}{\bibfnamefont{S.~H.} \bibnamefont{Parekh}},
  \bibnamefont{and} \bibinfo{author}{\bibfnamefont{D.~A.}
  \bibnamefont{Fletcher}}, \bibinfo{journal}{Nature}
  \textbf{\bibinfo{volume}{445}}, \bibinfo{pages}{295} (\bibinfo{year}{2007}).

\bibitem[{\citenamefont{Kempe et~al.}(2004{\natexlab{a}})\citenamefont{Kempe,
  Kornfield, and
  Lal}}]{2004_11_Macromolecules_Vol:37_Pg:8730--8738_Kempe.M;Kornfield.J;etal}
\bibinfo{author}{\bibfnamefont{M.~D.} \bibnamefont{Kempe}},
  \bibinfo{author}{\bibfnamefont{J.~A.} \bibnamefont{Kornfield}},
  \bibnamefont{and} \bibinfo{author}{\bibfnamefont{J.}~\bibnamefont{Lal}},
  \bibinfo{journal}{Macromolecules} \textbf{\bibinfo{volume}{37}},
  \bibinfo{pages}{8730} (\bibinfo{year}{2004}{\natexlab{a}}).

\bibitem[{\citenamefont{Kempe et~al.}(2004{\natexlab{b}})\citenamefont{Kempe,
  Kornfield, Ober, and
  Smith}}]{2004_05_Macromolecules_Vol:37_Pg:3569--3575_Kempe.M;Kornfield.J;eta%
l}
\bibinfo{author}{\bibfnamefont{M.~D.} \bibnamefont{Kempe}},
  \bibinfo{author}{\bibfnamefont{J.~A.} \bibnamefont{Kornfield}},
  \bibinfo{author}{\bibfnamefont{C.~K.} \bibnamefont{Ober}}, \bibnamefont{and}
  \bibinfo{author}{\bibfnamefont{S.~D.} \bibnamefont{Smith}},
  \bibinfo{journal}{Macromolecules} \textbf{\bibinfo{volume}{37}},
  \bibinfo{pages}{3569} (\bibinfo{year}{2004}{\natexlab{b}}).

\bibitem[{\citenamefont{Palffy-Muhoray and
  Meyer}(2004)}]{2004_03_Nature-Materials_Vol:3_Pg:139--140_Palffy-Muhoray.P;M%
eyer.R}
\bibinfo{author}{\bibfnamefont{P.}~\bibnamefont{Palffy-Muhoray}}
  \bibnamefont{and} \bibinfo{author}{\bibfnamefont{R.~B.} \bibnamefont{Meyer}},
  \bibinfo{journal}{Nature Materials} \textbf{\bibinfo{volume}{3}},
  \bibinfo{pages}{139} (\bibinfo{year}{2004}).

\bibitem[{\citenamefont{Kempe et~al.}(2004{\natexlab{c}})\citenamefont{Kempe,
  Scruggs, Verduzco, Lal, and
  Kornfield}}]{2004_03_Nature-Materials_Vol:3_Pg:177--182_Kempe.M;Scruggs.N;et%
al}
\bibinfo{author}{\bibfnamefont{M.~D.} \bibnamefont{Kempe}},
  \bibinfo{author}{\bibfnamefont{N.~R.} \bibnamefont{Scruggs}},
  \bibinfo{author}{\bibfnamefont{R.}~\bibnamefont{Verduzco}},
  \bibinfo{author}{\bibfnamefont{J.}~\bibnamefont{Lal}}, \bibnamefont{and}
  \bibinfo{author}{\bibfnamefont{J.~A.} \bibnamefont{Kornfield}},
  \bibinfo{journal}{Nature Materials} \textbf{\bibinfo{volume}{3}},
  \bibinfo{pages}{177} (\bibinfo{year}{2004}{\natexlab{c}}).

\end{thebibliography}

\end{document}